\documentclass[12pt]{article}
\begin{document}
\title{Comment on the paper of Jaffe et al (astro-ph/0503213) on evidence of vorticity  
and shear of the Universe}
\author{D. PALLE \\
Zavod za teorijsku fiziku \\
Institut Rugjer Bo\v skovi\'c \\
Po\v st. Pret. 180, HR-10002 Zagreb, CROATIA}
\date{ }
\maketitle

\vspace{5 mm}
{\it
We show that recent results in the paper of Jaffe et al on 
evidence of vorticity and shear of the Universe are incorrect.
They fit WMAP data with certain Bianchi model which
is known to have a vanishing vorticity.
The performed fit of the shear is not reliable because 
the respective formula seems not to be gauge invariant.
We suggest a theoretical framework to analyze asymmetric
large scale WMAP data.
}
\vspace{5 mm}

Few independent analyses of WMAP data claim the existence 
of the anisotropy in the large-scale spectrum of the CMBR 
fluctuations.
This stimulates a study of the data by cosmological models
beyond that of the Robertson-Walker.
Jaffe et al \cite{jaffe} attempt to fit data
with some Bianchi model using formulae of
Ref. \cite{barrow} with their definition of
"vorticity" and shear.
However, it is very well known that standardly
defined spacetime vorticity($\omega$)\cite{ehlers} vanishes
for these Bianchi models \cite{ellis}

\begin{eqnarray*}
\omega_{ab}=\omega_{[ab]}=h_{a}^{\ c}h_{b}^{\ d}
u_{[c;d]}, \\
\omega=[\frac{1}{2}\omega^{ab}\omega_{ab}]^{1/2}, \\
{[ab]}=\frac{1}{2}(ab-ba),\ u^{a}u_{a}=-1,\ h_{ab}=g_{ab}+u_{a}u_{b}.
\end{eqnarray*}

The apparent "vorticity" of Ref. \cite{barrow} is just
certain space rotation proportional to the shear components,
as can be seen from their formulae.
The standard spacetime vorticity is completely
independent physical observable, such as the shear, and it is a part 
of the decomposition of the covariant derivative of the velocity
vector \cite{ehlers}

\begin{eqnarray*}
u_{a;b}=\frac{1}{3}h_{ab}\Theta+
\sigma_{ab}+\omega_{ab}-a_{a}u_{b}, \\
\Theta=u^{a}_{\ ;a},\ a_{a}=u_{a;b}u^{b}, \\
\sigma_{ab}=h_{a}^{\ c}h_{b}^{\ d}(u_{(c;d)}
-\frac{1}{3}h_{cd}\Theta), \\
\sigma=[\frac{1}{2}\sigma_{ab}\sigma^{ab}]^{1/2},
\ (ab)=\frac{1}{2}(ab+ba).
\end{eqnarray*}

One cannot accept even the estimate of the shear($\sigma$) \cite{jaffe}
because they apply some relations for the fluctuations of
the CMBR \cite{barrow} which are not proved to be 
gauge invariant \cite{bardeen,bruni}.
We can hardly believe that one-year WMAP data are 
sensitive to such a small shear, estimated to be ten orders of magnitude 
smaller than the Hubble constant \cite{jaffe}.

Few years ago we proposed \cite{palle1}
a method to extract the shear by the Sachs-Wolfe effect
from very difficult measurements of quasar 
absorption spectra.

It is also possible to quantify vorticity  
in gauge invariant way, calculating the spatial gradients
of the CMBR fluctuations at large scales \cite{palle2}.
Complete all-scale analysis of the fluctuations requires the fit with 
 a gauge-invariant evolution of density contrasts in
models beyond that of the Robertson-Walker \cite{lasenby}.


\begin{thebibliography}{10}
\bibitem{jaffe} T. R. Jaffe et al: {\bf astro-ph/0503213}
\bibitem{barrow} J. D. Barrow, R. Juszkiewicz and D. H. Sonoda:
  MNRAS {\bf 213} (1985) 917
\bibitem{ehlers} J. Ehlers: Akad. Wiss. Lit. (Mainz),
  Abhandl. Math.-Nat. Kl. No. {\bf 11} (1961)
\bibitem{ellis} G. F. R. Ellis and M. A. H. MacCallum:
  Comm. Math. Phys. {\bf 12} (1969) 108
\bibitem{bardeen} J. M. Bardeen: Phys. Rev. {\bf D 22} (1980) 1882
\bibitem{bruni} G. F. R. Ellis and M. Bruni:
  Phys. Rev. {\bf D 40} (1989) 1804
\bibitem{palle1} D. Palle: Nuovo Cimento {\bf 116 B} (2001) 1317
\bibitem{palle2} D. Palle: {\bf astro-ph/0407122}, Nuovo Cimento (2005) in press
\bibitem{lasenby} A. D. Challinor and A. N. Lasenby: Phys. Rev. {\bf D 58}
  (1998) 023001; A. D. Challinor and A. N. Lasenby: ApJ {\bf 513} (1999) 1;
  T. Gebbie and G. F. R. Ellis: Ann. Phys. {\bf 82} (2000) 285
\end{thebibliography}
\end{document}